\documentclass[
    aps,
    prb,
    showpacs,
    reprint,
    superscriptaddress,
    longbibliography,
    floatfix,
    amsmath
]{revtex4-1}

\pdfoutput=1
\usepackage{braket}
\usepackage{graphics}
\usepackage{graphicx}
\usepackage{multirow}
\usepackage{tasks}
\usepackage{dcolumn}
\usepackage{scrextend}
\usepackage{longtable}
\usepackage{dblfnote}
\usepackage[multiple]{footmisc}
\newcolumntype{.}{D{.}{.}{-1}}

\usepackage{epsfig}
\usepackage{epstopdf}
\usepackage{hyperref}
\usepackage{verbatim}
\usepackage[english]{babel}

\newcommand{\NMaterials}{35 }

\begin{document}

    \title{Fast and accessible first-principles calculations of vibrational properties of materials}
    
    \author{Timur Bazhirov and E. X. Abot}
    \thanks{mail-to: info@exabyte.io}
    \affiliation{Exabyte Inc., San Francisco, California 94103, USA}

    \begin{abstract}

    We present example applications of an approach to first-principles calculations of vibrational properties of materials implemented within the Exabyte platform\cite{exabytePlatform, 2018-exabyte-accessible-CMD, 2018-exabyte-binary-compounds}. We deploy models based on the Density Functional Perturbation Theory to extract the phonon dispersion relations and densities of states for an example set of \NMaterials samples and find the results to be in agreement with prior similar calculations. We construct modeling workflows that are both accessible, accurate, and efficient with respect to the human time involved. This is achieved through efficient parallelization of the tasks for the individual vibrational modes. We report achieved speedups in the 10-100 range, approximately, and maximum attainable speedups in the 30-300 range, correspondingly. We analyze the execution times on the current up-to-date computational infrastructure centrally available from a public cloud provider. Results and all associated data, including the materials and simulation workflows, are made available online in an accessible, repeatable and extensible setting.

\end{abstract}
    \maketitle
    \section{Introduction}
\label{sec:introduction}

    Properties associated with the dynamics of a crystal lattice are of key importance for the design and discovery of new materials from nanoscale. Such properties can be computationally obtained by studying phonons - collective excitation in a periodic arrangement of atoms characterized by the modes of vibration and the corresponding frequencies. The phonon dispersion relations facilitate the calculation of many transport properties: heat capacity and vibrational contribution to the entropy of the system, thermal conductivity, superconductivity, and ferroelectricity\cite{petretto2018materialsproject-phonons}. Phonon spectra also provide the information about the phase stability of compounds through the inspection of imaginary phonon modes and facilitate the interpretation of Raman spectra\cite{petretto2018materialsproject-phonons}. 
    
    Data-driven approaches rooted in the first-principles modeling techniques recently received much attention for the purpose of the design and discovery of new materials, with multiple success stories reported to date\cite{jain2013materialsproject, curtarolo2012aflowlib, saal2013openQMD, pizzi2016aiida, nomad}. First-principles computational techniques have also long been known as an accurate and reliable way to predict the vibrational properties of materials\cite{diMeo2009QEphononsGrid, QE2009mainReference}. Although the efforts in organizing the calculation data have also existed for some time\cite{phonopy2015}, due to the associated computational complexity a high-throughput approach to the calculations of the vibrational properties of materials was reported only recently\cite{petretto2018materialsproject-phonons, AFLOW2017thermalCond, petretto2017DFPTConvergence}.
    
    Density Functional Perturbation Theory\cite{gonze1996DFPT} is a first-principles technique that allows one to extract the information about how a material responds to an atomic vibration at a specific frequency and shape. This approach is well established and accurate\cite{QE2009mainReference, diMeo2009QEphononsGrid}, however, computationally demanding due to having to repeat the total energy calculations for each of the perturbation shapes of the original crystal lattice. For a unit cell with N atoms, the total number of phonon modes is 3N per each perturbation. Since multiple perturbations need to be sampled to achieve accuracy, the complexity of phonon calculations often is two or more orders of magnitude higher than for the total energy calculations\cite{petretto2017DFPTConvergence}.
    
    We present example applications of the approach implemented inside Exabyte platform\cite{exabytePlatform} and similar to previously described in Refs. \cite{2018-exabyte-accessible-CMD, 2018-exabyte-binary-compounds}. The approach is able to facilitate high-throughput first-principles calculations in a repeatable way transferable from one material to another. In this manuscript we apply this approach for the extraction of the vibration properties of materials. We use Density Functional Perturbation Theory in the plane-wave pseudopotential approximation\cite{hohenberg-kohn1964DFT, mlcohen1979pseudopotentialDFT, gonze1996DFPT, QE2009mainReference} and obtain the phonon dispersions and densities of states for a set of \NMaterials materials. We optimize the modeling workflows in order to minimize the human time required per each calculation, and obtain the frequencies per each individual irreducible representation of the phonon perturbation in parallel\cite{diMeo2009QEphononsGrid}. We compare the results with the available reference data from other authors and find good agreement.
    
    This manuscript is structured as follows. We first explain the materials studied and discuss the logic, the methodology, and the parameters used inside the calculation workflows. Next, we present example results for a control subset of materials and compare them with the available computational data from other authors. Finally, we discuss the results and the achieved speedups more in depth and suggest the pathways toward further improvements. This work presents all the following: the results, the tools that generated the results, all associated data, and an easy-to-access way to reproduce and further improve upon our work.\cite{exabytePlatformPHURL}

\section{Methodology}
\label{sec:methodology}

\subsection{General logic}

    We demonstrate the general execution flow employed in this work in Ref. \cite{2018-exabyte-accessible-CMD}. We employ the framework explained therein in order to construct the simulation workflows for the calculation of the vibrational properties discussed here. The users of Exabyte platform can clone the associated entities (eg. materials, workflows) - and re-create our calculations in order to reproduce or further improve the results.

\subsection{Materials}

    All materials studied in this work constitute a subset of the ESC-71 set from \cite{2018-exabyte-accessible-CMD} with the total count of \NMaterials. The details about the materials studied are given in Fig \ref{table:materials}. Our selection is based initially on the widely used semiconducting compounds that have relatively small crystal unit cells.

    \begin{table}[h!]
    	\centering
    	\begin{tabular}{l  l  c  |  l  l  c }
    		\hline
    		\hline
            Formula & MP id & $N_{at}$ & Formula & MP id & $N_{at}$ \\
            \hline
            \hline
            Si      &  mp-149     & 2 & Li2O    &  mp-1960    & 3 \\
            Ge      &  mp-32      & 2 & BN      &  mp-7991    & 4 \\
            Bi      &  mp-23152   & 2 & AlN     &  mp-661     & 4 \\
            Sn      &  mp-117     & 2 & CaO     &  mp-545512  & 4 \\
            BP      &  mp-1479    & 2 & MgSe    &  mp-1018040 & 4 \\
            GaP     &  mp-2490    & 2 & MgTe    &  mp-1039    & 4 \\
            AlAs    &  mp-2172    & 2 & SnO2    &  mp-856     & 6 \\
            GaAs    &  mp-2534    & 2 & AlGaAs  &  N/A        & 8 \\
            GaN     &  mp-830     & 2 & InGaAs  &  N/A        & 8 \\
            YN      &  mp-2114    & 2 & InGaP   &  N/A        & 8 \\
            ZnS     &  mp-10695   & 2 & AlInAs  &  N/A        & 8 \\
            BeSe    &  mp-1541    & 2 & AlInSb  &  N/A        & 8 \\
            MgO     &  mp-1265    & 2 & GaAsP   &  N/A        & 8 \\
            MgS     &  mp-1315    & 2 & AlGaN   &  mp-1019508 & 8 \\
            ZnO     &  mp-2229    & 2 & B2O3    &  mp-717     & 10 \\
            BeO     &  mp-1778    & 2 & Al2O3   &  mp-1143    & 10 \\
            BaSe    &  mp-1253    & 2 & B       &  mp-160     & 12 \\
            CaSe    &  mp-1415    & 2 \\
            \hline
            \hline
    	\end{tabular}
    	\caption{
    	    Materials studied in this work with the corresponding identificators from Ref. \cite{jain2013materialsproject}. $N_{at}$ - number of sites (atoms) in the crystal unit cell. Consult Ref. \cite{2018-exabyte-accessible-CMD} for additional details.
    	    }
    	\label{table:materials}
    \end{table}

\subsection{Workflows}

\begin{figure*}[ht!]
        \label{fig:workflow}
        \centering
        \includegraphics[width = 0.95\textwidth]{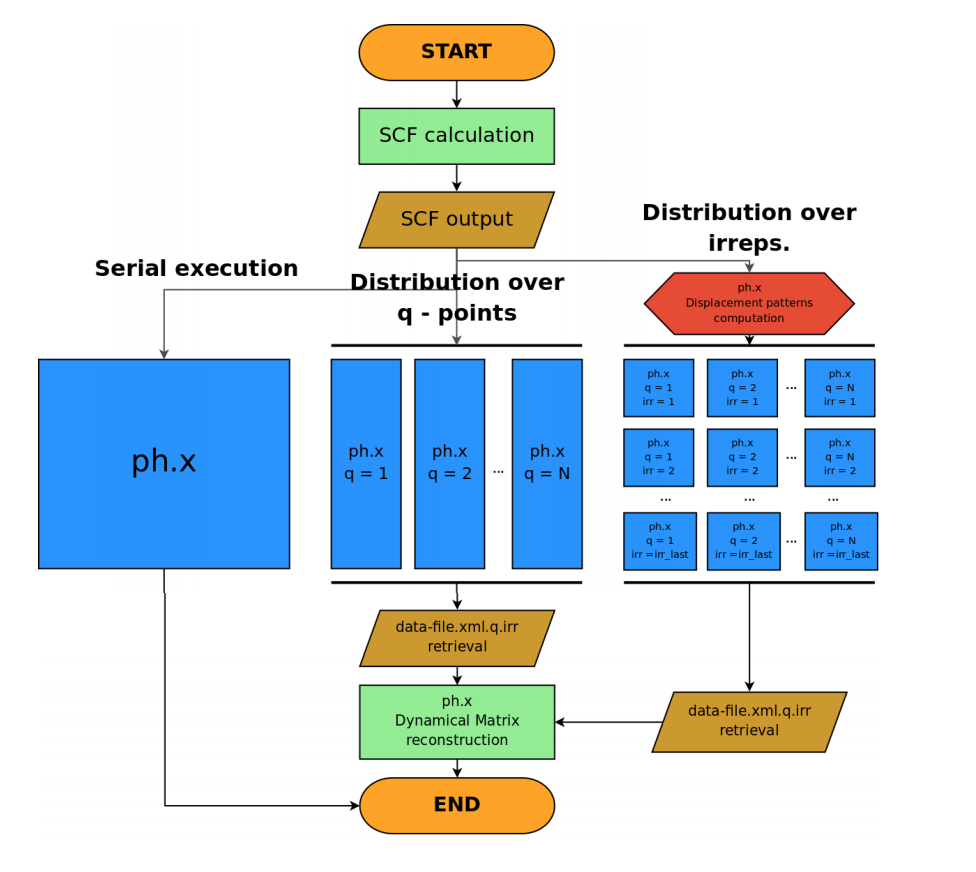}
        \caption{
            Flowchart depicting the different approaches to the calculation of the phonon dispersions as implemented in Ref. \cite{QE2009mainReference}. The approach used in this work is the rightmost one. "SCF" stands for the self consistent field calculation. "ph.x" denotes the phonon calculations by means of Density Functional Perturbation Theory. "irrep" is an irreducible representation of a vibrational mode. Reproduced from Ref. \cite{diMeo2009QEphononsGrid}.
        }
    \end{figure*}

\begin{figure*}[ht!]
        \label{fig:workflow-platform}
        \centering
        \frame{\includegraphics[width = 0.95\textwidth]{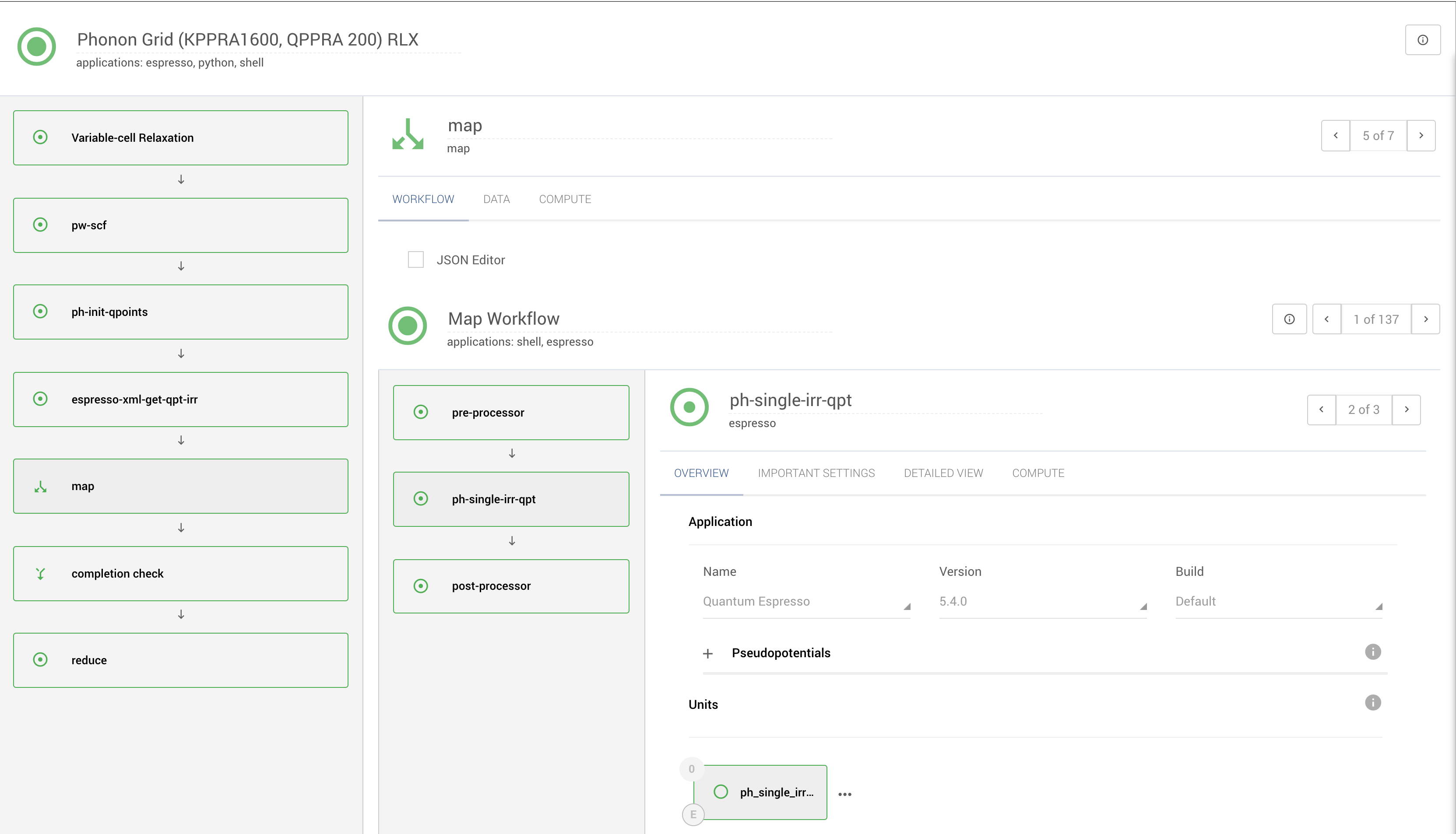}}
        \caption{
            Example representation of the grid-parallel (or "map-reduce") phonon calculation workflow in Exabyte platform. The left panel contains multiple sub-workflows. The "map" step containing a pre-processor, phonon calculation for a single irreducible representation, and a post-processor is expanded in the right panel. As it can be seen, a total of 137 independent calculations is performed. Web page with this data is available at the link in Ref. \cite{exabytePlatformPHURL}.
        }
    \end{figure*}

    We implement a grid-parallel workflow for the calculation of the phonon dynamical matrices initially explained in Ref. \cite{diMeo2009QEphononsGrid} and demonstrated in Figs. \ref{fig:workflow} and \ref{fig:workflow-platform} with an additional optional preceeding step for a variable-cell relaxation. During the phonon calculation part the following happens:
    
    \begin{enumerate}
        \item First, the irreducible representations for the vibrational modes (irreps) are generated based on the sampling grid in the reciprocal space (q-point grid). Full symmetry analysis is not performed in the current implementation.
        \item Second, a separate calculation is prepared and submitted for execution to the cloud infrastructure manager per each irreducible representation ("map" stage).
        \item Next, the computational infrastructure is provisioned on-demand at a cloud provider with a cap on the total number of nodes as explaned further in this section.
        \item Finally, after the calculations for all irreps are finished, the dynamical matrices are collected and phonon dispersions and density of states are calculated ("reduce" stage).
    \end{enumerate}
    
    Thus, we employ a "map-reduce" type embarassingly parallel scenario, and couple the calculations of the individual phonon dispersion modes with the allocation of computational resources on the cloud. This allows for the improved efficiency and speedup, such that the limiting phase in the total calculation is the longest run per individual irreducible representation. Unlike the previously considered workflows categorization based on the inclusion of the semi-core states, spin-orbit coupling and magnetism \cite{2018-exabyte-accessible-CMD}, in this work we omit the considerations of the latter two and only include semi-cores as it is implemented in the GBRV pseudopotential set\cite{GBRV2013pseudopotentials}.

\subsection{Computational setup}

    We use Density Functional Theory\cite{kohn1965self} in the planewave pseudopotential formalism\cite{mlcohen1979pseudopotentialDFT} as implemented in Quantum ESPRESSO (QE) package\cite{QE2009mainReference}. Within the generalized gradient approximation the exchange-correlation effects were modeled using the Perdew-Berke-Ernzerhof (PBE)\cite{perdew1996generalized} functional. The ultrasoft GBRV pseudopotentials at version 1.5\cite{GBRV2013pseudopotentials} with the recommended cutoff values of 40 Ry and 200 Ry are used for the electronic wavefunctions and electronic densities correspondingly. We implemented sampling in the reciprocal cell based on k-points per reciprocal atom (KPPRA) with a uniform unshifted grid. A minimum KPPRA of 1,600 was used for the electronic structure calculations. The phonon properties are calculated on an grid that corresponds to a minimum QPPRA of 200. The Fourier transform and subsequent interpolation to an effective IPPRA (interpolated points per reciprocal atom) of 12,800 was used, as it is implemented in QE through "q2r" and "matdyn" modules.
    
    All calculations were performed using the hardware available from Microsoft Azure cloud computing service\cite{azure-instance-types} in the same manner as described in Refs. \cite{2018-exabyte-accessible-CMD, 2018-exabyte-binary-compounds}, except for the use of "F16" instances for the current work. Computational resources were provisioned and assembled on-demand by software implemented and available within the Exabyte platform\cite{exabytePlatform}. All runs were executed by a single person within a one-week period in June 2018, which emphasizes the power of the underlying general approach to materials modeling implemented in Exabyte platform. The peak size of the computational infrastructure used during this work was administratively limited to 200 nodes or 3,200 total computing cores

\subsection{Data access and repeatability}

\begin{figure}[t!]
        \label{fig:results-ingaas}
        \centering
        \frame{\includegraphics[width = 0.48\textwidth]{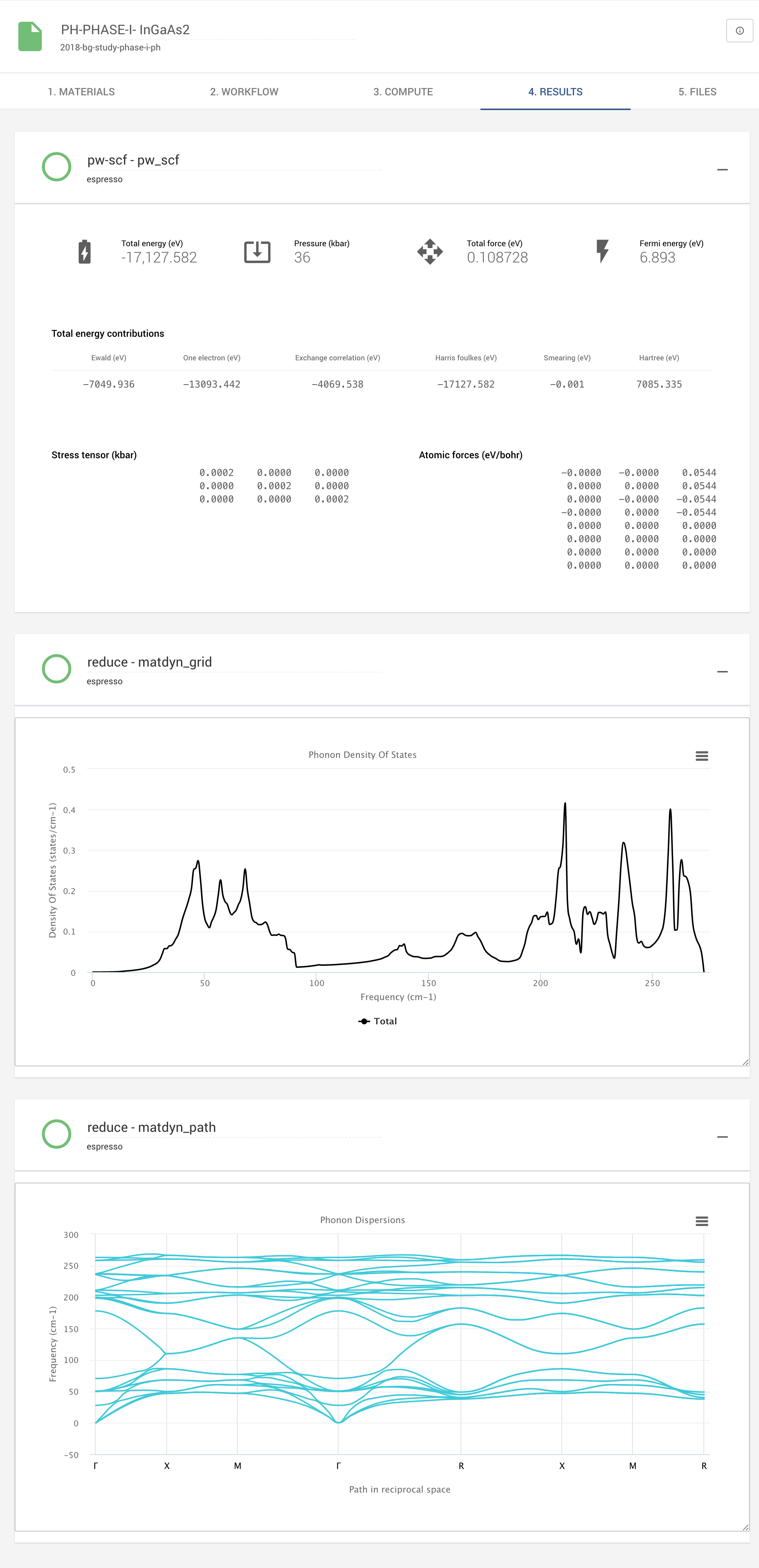}}
        \caption{
            Example results for an 8-atom InGaAs cell, as available within the Exabyte platform\cite{exabytePlatformPHURL}. The top panel lists parameters for the initial self-consistent field calculation, and the bottom two contain the plots for the phonon density of states and the phonon dispersions.
        }
    \end{figure}

\begin{figure}[h!]
        \label{fig:results-b}
        \centering
        \frame{\includegraphics[width = 0.38\textwidth]{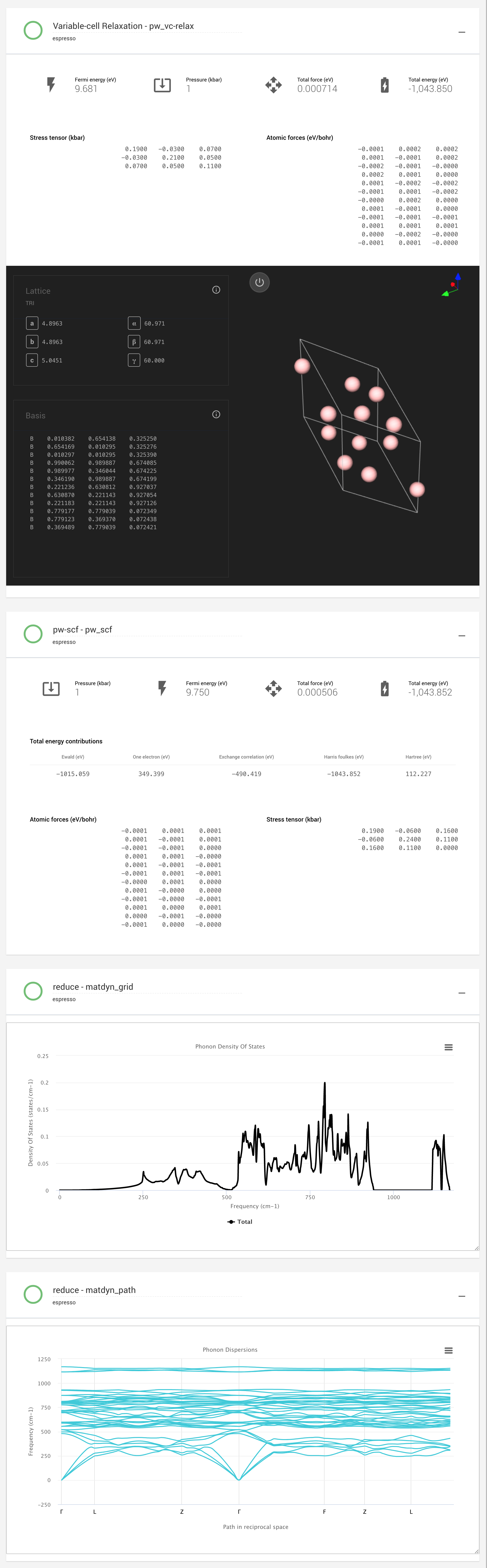}}
        \caption{
            Results page for a 12-atom boron calculation.
        }
    \end{figure}

    The materials, workflows, batch jobs for each material with the associated properties, and files for each step of the simulation workflows are all freely available online at the link in Ref.\cite{exabytePlatformPHURL}. Readers interested in repeating or imporving upon our work may create an account, copy materials and/or workflows to their account collection, and recreate the simulation for this materials. An example simulation workflow, as employed in the current work, is presented in Fig. \ref{fig:workflow-platform}. Readers can see the map-reduce type logic included in the workflow, where the individual calculation tasks are performed independently in parallel in order to speed up the execution. Example results for InGaAs and B, in the same exact representation as can be accessed through Ref. \cite{exabytePlatformPHURL}, are shown in Fig. \ref{fig:results-ingaas} and Fig. \ref{fig:results-b} correspondingly.

    \section{Results and Discussion}
\label{sec:results}
    
    \subsection{Results and comparison with prior calculations}
    \label{subsec:results}
    
\begin{figure*}[ht!]
        \vspace{0.5cm}
        \centering
        \includegraphics[width = 0.99\textwidth]{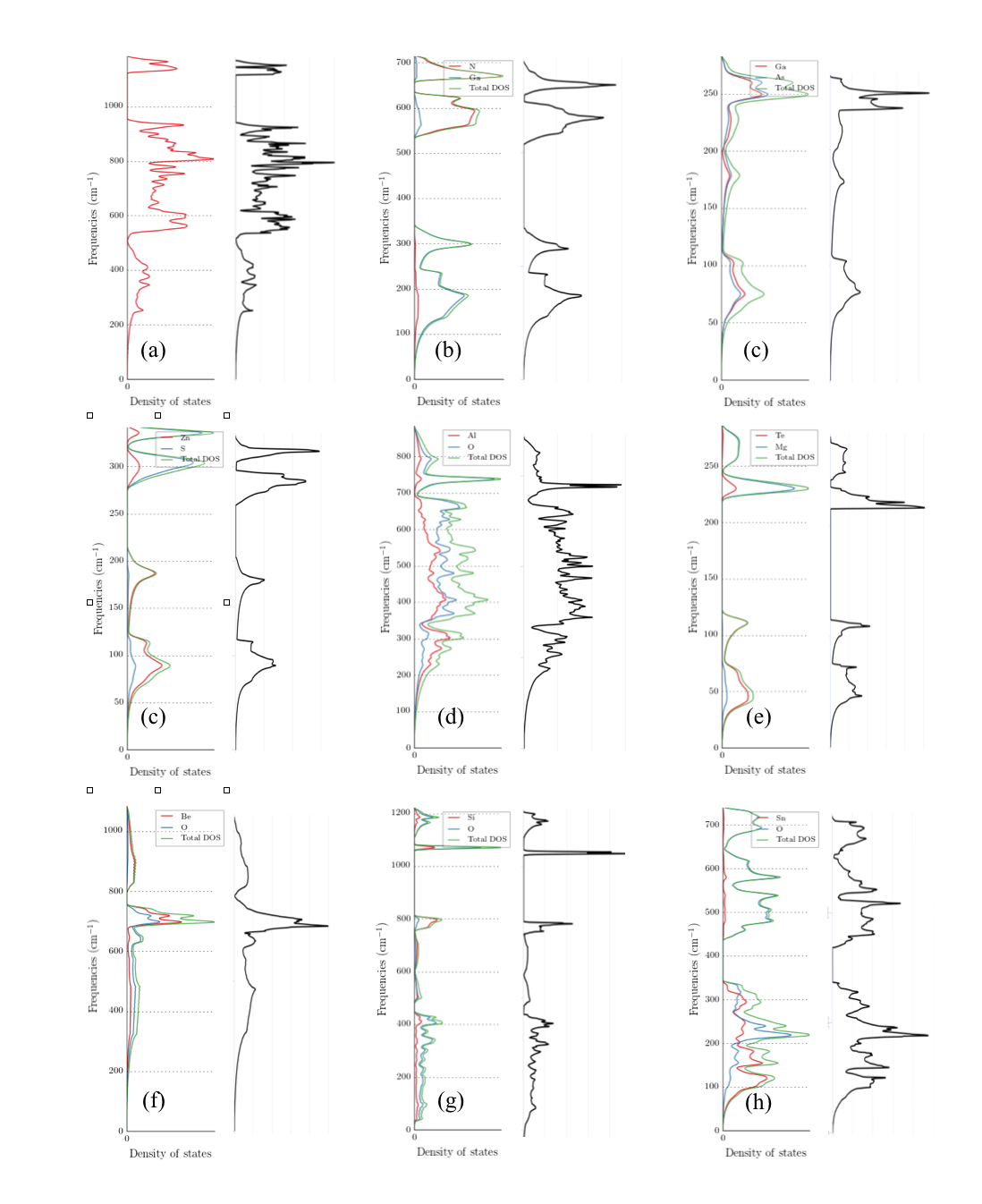}
        \vspace{0.5cm}
        \caption{
            Comparative plot of the results for the phonon dispersions calculated in this work and available in MP database\cite{jain2013materialsproject}. (a) - boron, (b) - GaN, (c) - GaAs, (d) - ZnS, (f) - sapphire Al$_2$O$_3$, (e) - MgTe, (f) - BeO, (g) - SiO$_2$, (h) - SnO$_2$ ( Left panels for each sub-figure show MP data and contain the projected density of states for each atom. Right sub-figure panels has the total phonon density of states calculated in this work.
        }
        \label{fig-all}
    \end{figure*}

    The results for all the materials studied in this work are available online at the link in Ref. \cite{exabytePlatformPHURL}.
    Fig. \ref{fig-all} shows a comparison of the calculated phonon density of states for a subset of 9 compounds, including boron, GaN, GaAs, ZnS, Al$_2$O$_3$, MgTe, BeO, SiO$_2$, SnO$_2$, with the results of the Materials Project\cite{jain2013materialsproject} (further referred to as MP). As it can be seen, the results are in agreement with each other in the overall shape, with a small (1-5\%) shift toward the high frequency range in MP case. We attribute this shift to the use of PBEsol\cite{pbeSOL2008perdew} functional in their work, versus PBE\cite{perdew1996generalized} for our calculations.
    
    \subsection{Goals}
    \label{subsec:goals}

    We meant this study as a demonstration of the capabilities of Exabyte platform in deploying the density functional perturbation theory (DFPT) tools to predict the vibrational properties of materials. We also focused our attention on how it can be accelerated and applied in an accessible way with minimal additional computational setup (i.e. no specialized hardware or compilation routines). We elaborate on the results of our prior work\cite{2018-exabyte-accessible-CMD} and extend the spectrum of materials properties available.
    
    \subsection{Further improvements}
    \label{subsec:improvements}

    We can consider multiple ways to further improve the results presented in this work. Firstly, as mentioned in \cite{petretto2018materialsproject-phonons}, employing PSEsol instead of the more widely used (at the moment of this writing) PBE functional for the treatment of the exchange-correlation in the materials studied might be beneficial. According to the comparison presented in Fig. \ref{fig-all} this effect is expected to be small - within 1-5\%. Secondly, a rigorous relaxation routine, together with a more extensive convergence study for the reciprocal cell sampling in each of the computed cases might be beneficial. As the magnitude of the phonon frequencies is relatively small (meV range), the artifacts related to convergence can sometimes lead to the artificially present negative frequencies\cite{petretto2017DFPTConvergence}. We originally attempted to include all materials from the ESC-71 set\cite{2018-exabyte-accessible-CMD}, and left some out due to the time constraints related to resolving the presence of negative frequencies. Lastly, an improvement in speed and efficiency of the modeling workflows may be achieved by further optimizing the coupling of the computational infrastructure to the individual calculations per each irreducible representation. As the Table \ref{table:runtimes} demonstrates, the minimum attainable runtime with the "map-reduce" workflow can be as much as 374 shorter than for a sequential, while we practically achieved speedups up to 134.
    
    \subsection{Computational time and cost}
    
    \begin{table}[ht!]
        \label{table:runtimes}
    	\centering
    	\begin{tabular}{l | c |  c   l  c | c}
    		\hline
    		\hline
    		Material & $N_{at}$  & $t_{seq}$, (hr) & $t_{seq}/t_{act}$ & $t_{seq}/t_{min}$ & Cost ($\$$) \\
    	    \hline
    	    \hline
      	    MgO      & 2      & 44.1    & 134  & 174  &  100   \\
      	    MgSe     & 4      & 66.3    & 30   & 248  &  150   \\
      	    AlGaN    & 8      & 18.6    & 13   & 34   &  50   \\
      	    B        & 12     & 102.3   & 117  & 374  &  250   \\
    		\hline
    		\hline
    	\end{tabular}
    	\caption{
    	    Average calculation times for a selected subset of materials. $N_{at}$ - number of sites (atoms) in the crystal unit cell. $t_{seq}$ - phonon calculation runtime for a sequential mode, where all irreducible representations are calculated one-by-one (leftmost branch in Fig. \ref{fig:workflow}). $t_{act}$ - actual runtime as recorded in Exabyte platform. $t_{min}$ - minimum possible runtime for the phonon part, equal to the maximum calculation time per a single irreducible representation. Approximate costs are given as well.
    	}
    \end{table}    
    
    We present the analysis of the runtimes for the different calculation scenarios. First, we list the total "sequential" execution time $t_{seq}$ that a phonon calculation would take without parallelizing the tasks for irreducible representations. In practice, this would mean confining the simulation to a single computing node with 16 cores as explained in section \ref{sec:methodology}. Next, we present the corresponding speedup ratios for the actual runtime recorded $t_{act}$ and for the minimum attainable runtime $t_{min}$ corresponding to the longest run among all irreps. As can be seen from the table, we achieved speedups in the 13-134 range in practice, while corresponding maximum speedups possible are 34-374. We also present the associated costs, which, due to the elastic nature of cloud computing, do not depend on the specific calculation scenario.

    \subsection{Future outlook}

    The landscape of computational materials design is rapidly evolving toward a data-driven science, with multiple initiatives contributing toward the automated aggregation and categorization of materials properties. Major improvements in the way computational materials science is used would be possible when the range of materials properties feasible for calculation is extended to include the vibrational spectra and related. The approach described in this work can assist with the above. This work demonstrates that high-fidelity data about the vibrational properties, perhaps only for the electronic materials at this moment, is readily attainable in an accessible and repeatable manner. Our intent is to welcome collaborative contributions in order to, firstly, further grow the online repository of the results; secondly, allow contributions from other modeling techniques beyond studied here; and, finally, facilitate the creation of statistical (machine learning) models based on the available data.

    \section{Conclusions}
\label{sec:conclusions}
    
    We present the applications of a novel approach to materials modeling from nanoscale implemented within the Exabyte platform\cite{exabytePlatform} and capable of rapidly delivering results about the vibrational properties of materials in an accessible and data-centric manner. We apply this approach to a set of \NMaterials materials in order to demonstrate how it works. We report the results for the phonon densities of states and phonon dispersions obtained using the Density Functional Perturbation Theory within the Generalized Gradient Approximation (GGA). We compare the results with prior similar calculation attempts and discuss the corresponding computational costs and pathways to further improvements.
    
    We demonstrate how computationally demanding task of calculating the phonon frequencies, that would otherwise take from 18 to 102 hours on an up-to-date high-performance computing server, can be accelerated by a factor of 13-134 such that the resulting runtime fits within one hour. We present not only the results and the associated data, but also an easy-to-access way to reproduce and extend the results by means of the Exabyte platform.\cite{exabytePlatformPHURL} Our work provides an accessible and repeatable practical recipe for performing high-fidelity first-principles calculations of the vibrational properties of materials in a high-throughput manner.

    \bibliography{references}

\end{document}